# Overcoming the absorption limit in high-harmonic generation from crystals

Hanzhe Liu[1,2*], Giulio Vampa[1*], Jingyuan Linda Zhang[3,4], Yu Shi[4,5], Siddharth Buddhiraju[4,5], Shanhui Fan[3,4,5], Jelena Vuckovic[3,4,5], Philip H. Bucksbaum[1,2,3,6], David A. Reis[1,3,6]

[1] Stanford PULSE Institute, SLAC National Accelerator Laboratory, Menlo Park, California 94025, USA

[2] Department of Physics, Stanford University, Stanford, California 94305, USA

[3] Department of Applied Physics, Stanford University, Stanford, California 94305, USA

[4] E. L. Ginzton Laboratory, Stanford University, Stanford, California 94305, USA

[5] Department of Electrical Engineering, Stanford University, Stanford, California 94305, USA

[6] Department of Photon Science, Stanford University, Stanford, California 94305, USA

* These authors contributed equally to this work

**Since the new millennium coherent extreme ultra-violet and soft x-ray radiation has revolutionized the understanding of dynamical physical[1], chemical[2] and biological systems[3] at the electron's natural timescale. Unfortunately, coherent laser-based upconversion of infrared photons to vacuum-ultraviolet and soft x-ray high-order harmonics in gaseous[4-6], liquid[7] and solid[8-12] targets is notoriously inefficient[13,14]. In dense nonlinear media, the limiting factor is strong re-absorption of the generated high-energy photons[13,14]. Here we overcome this limitation by allowing high-order harmonics generated from a periodic array of thin one-dimensional crystalline silicon ridge waveguides to propagate in the vacuum gaps between the ridges, thereby avoiding the high absorption loss of the bulk nonlinear material and resulting in a ~ 100-fold increase in propagation length. As the grating period is varied, each high-harmonic shows a different and marked modulation, indicating the onset of coherent addition which is otherwise suppressed in absorption-limited emission[13,15]. By beating the absorption limit, our results pave the way for bright coherent short-wavelength sources and their implementation in nano-photonic devices.**

In attosecond photonics, the oscillating electric field of light drives numerous coherent processes in gases[1], liquids[7] and solids[8-12]. Solids provide a suitable platform for technological applications of these processes, such as for petahertz electronics[16-18], electric-field sensing[19] and on-chip extreme ultra-violet microscopy. The latter requires efficient generation of high-order harmonics at the nanoscale and low-loss transport on a chip. Unfortunately, due to strong reabsorption of above-bandgap high-harmonics, propagation of high-energy photons in solids is essentially forbidden[15], limiting both the flux buildup and on-chip integration.

Here we report on a suitably designed nano-patterned crystal that overcomes the intrinsic absorption limit of the bulk. We fabricate a periodic array of thin single-crystal silicon ridge-waveguides that spatially separate the nonlinear high-harmonic polarization, which travels along the ridge's sidewalls, from the generated high-harmonic waves, which leak and propagate inside the vacuum slots between ridges. The working principle is sketched in Fig. 1. The width of individual silicon ridge-waveguides is designed to be sub-wavelength in order to maintain a relatively high field strength on the ridges' sidewalls[20]. Generating high harmonics in the ridge-vacuum interface region (step 1) allows their coupling into the vacuum slots (step 2) and their propagation to the output of the crystal (step 3). This approach exploits the desired high density of crystals and the minimal loss of the vacuum channels, which constitute leaky waveguides for the high harmonics. Our main finding is that, as the slot width is varied, each deep- and vacuum-ultraviolet high harmonic shows strong modulations unique to each harmonic, indicating the guiding of the generated short-wavelength radiation in the vacuum channels, as well as the onset of coherent addition on otherwise absorption-limited radiation. Previous reports of enhanced high-harmonic emission from solids exploited nanoscale confinement of the infrared field[21-24]. Our structures exploit both the increased surface area around each waveguide and the reduced absorption and therefore results in a high-harmonic flux that surpasses that measured from uniform bulk.

In the experiment (Fig. 2**a**), we illuminate linearly tapered gratings consisting of 1 μm-tall Si ridges of continuously varying width and pitch with TM-polarized (magnetic field pointing parallel to the ridges) mid-infrared few-cycle femtosecond pulses with a center wavelength of

2.35 µm at a repetition rate of 80 MHz [25]. The estimated peak intensity in vacuum is ~ 0.2 TW/cm$^2$. The generated high-harmonics are collected with a Fresnel zone plate (FZP) and their spectrum is recorded with a vacuum ultraviolet (VUV) spectrometer and a silicon CCD camera (see Methods for details on the experimental setup and fabrication). The taper of the gratings is sufficiently small (the slot width varies by ~ 0.2 nm/µm) that the geometry variation within the laser focus is negligible. Varying the grating geometry by translating the structures along the taper (laterally with respect to the laser focus) results in variations of the high-harmonic spectrum, as shown in Fig. 2**b** for illumination of two sample regions with narrow (blue) and wide (red) slot widths. Fig. 2**c** shows a high-resolution image of the end of the gratings, on the wide slot side. Finely scanning the slot and Si ridge widths results in strongly modulated high-harmonic powers, as reported in Figs. 2**d-g**, blue lines, for harmonics 9$^{th}$ to 15$^{th}$. The grey-shaded areas result from illumination of bulk Si. The high-harmonic flux is measured relative to the maximum obtained from these bulk regions (corresponding to infrared powers close to the damage threshold). Each high harmonic shows a marked modulation with slot width which is unique to each harmonic order. The observed harmonic-order dependent, non-monotonic modulation, especially the sharp rise in power when the slot width is comparable to the harmonic wavelength, marks the guiding of high harmonics and the onset of propagation effects over distances much longer than the intrinsic propagation length in bulk Si (on the order of one absorption length, ~ 10 nm). This is our major experimental finding that corroborates the working principle presented in Fig. 1. At the same time, the simulated intensity of the infrared field on the Si ridges' sidewalls (yellow lines in Fig. 2**d**) exhibits a monotonic decrease as the slot width increases. This would result in a simultaneous decrease of all harmonics without the vacuum propagation effects. Additional arguments that indicate the order-dependent modulation is due to propagation effects rather than to variations of the excitation field strength and polarization can be made based on the measured power scaling of high harmonics and their dependence on the ellipticity of the infrared field. This discussion is included in the Supplementary Information.

More insight into the emission mechanism can be drawn from a quantitative analysis of the modulated high-harmonic power. Combining finite-difference frequency-domain (FDFD) simulations of the infrared driver with analytical propagation of harmonic waves in the leaky slot (see Supplementary Information for details about the theoretical framework), we predict the high-harmonic power reported in Fig. 2**d-g**, red lines. The expected power exhibits increasingly strong oscillations as the slot widens. The increasing amplitude results from a sharp increase in transmission of harmonic waves in the slot for slot widths approaching one harmonic wavelength. Our calculation reveals that the absorption length increases 100-fold, from < 20 nm to ~ 2 μm, as the blue lines in Fig. 2**h-k** show. The sharp increase in transmission of a slot is consistent with the maximum measured high-harmonic flux occurring for slot widths comparable or larger than one harmonic wavelength.

With increased transmission, phase mismatch between the harmonic wave and the nonlinear polarization propagating along the ridge sidewall leads to power oscillations with a period equal to one coherence length (red lines in Figs. 2**h-k** show the predicted coherence length). These are less visible in the experiment, likely due to imperfect experimental geometry, such as bowing of the ridges along the waveguide height, as shown in Fig. 2**c**; and due to frequency averaging over the infrared spectrum. Towards the wide slot end of the grating, the measured harmonic power decreases for all reported orders, in contradiction to the expected rise of another oscillation cycle. This is likely due to the monotonic decrease in infrared field strength (Fig.2**d**, yellow line), as the slot width increases, below the threshold for generating harmonics at the ridges' sidewalls. This strong-field threshold, a characteristic of all high-harmonic sources, is not included in the theoretical model.

An immediate advantage of exploiting the high transmission of the slots is that the high-harmonic flux can potentially outperform that from bulk crystals. Figure 3 shows the high-harmonic yield as a function of infrared power, until damage occurs, indicated by a sudden drop in the yield. Despite phase mismatch and the ~ 3 times less nonlinear-material, the harmonic yield from the device is comparable to that generated from bulk. In addition, Supplementary Figure 5 shows that narrower Si ridges and wider slots result in even stronger emission, likely due to a combination of the higher infrared field on the ridges' sidewalls and the increased transmission of the wider slots.

Breaking the absorption limit further implies the possibility of phase matching high-harmonic emission with dispersion-engineered structures[26], which could lead to significantly increased flux from nanoscale sources of high-harmonics. Figure 4 shows the estimated high-harmonic power for one grating geometry as a function of waveguide length, in the phase mismatch (red line) and phase matched (blue line) cases. While the former shows oscillations every coherence length, the latter shows an exponential increase ultimately only limited by the absorption length of the slot. A gain of ~ 100 is expected in the phase-matched case. A new design for on-chip integrated waveguides will allow patterning along the propagation direction, thus opening the possibility of tailoring the dispersion with sub-infrared-wavelength structures[26,27] and photonic crystals[28]. Quasi-phase matching schemes can potentially be employed, too[29]. Adiabatic phase matching also promises to convert infrared to high-harmonic power with near unity efficiency[30]. Wide bandgap crystals more suitable for emission of Extreme Ultra-violet wavelengths can potentially replace Si. For example, lithium niobate is already a well-developed nanophotonics platform[31]. These efforts will lead to high-flux short-wavelength sources confined to only tens or one hundred nanometers, which will be ideal for nanoscale high-harmonic microscopy.

To conclude, we have overcome one of the major obstacles to high-flux solid-state high-harmonic sources: the extremely short absorption length of bulk solids at deep- and vacuum-ultraviolet wavelengths (~ 10 nm). We have done so by generating high harmonics at the surface of a nano-structured nonlinear crystal, allowing the harmonic waves to leak and propagate in vacuum channels over micron-scale distances – a ~100-fold increase in transmission length. Our results suggest that it will be possible to extend nanophotonic devices that rely on propagating waves for applications and studies that benefit from short-wavelength radiation, such as on-chip chemically-sensitive spectroscopy and nanoscale imaging. Short-wavelength on-chip laser oscillators and frequency-combs can potentially be realized too, for example with slot-waveguide-based ring resonators[32]. Finally, the same concepts can be applied to liquid- and gas-phase high-harmonic generation, which also suffer from the small absorption length[7,14], at even shorter wavelengths. Micro-to-nanoscale patterning of liquids can be achieved with the help of microfluidic chips[33], and gas volumes can potentially be manipulated too if adsorbed on patterned surfaces. Therefore, we expect our findings have a potentially broad impact on optimizing a wide range of coherent short-wavelength sources.

## Methods

**Experiment Setup**

The laser source used in this experiment is a mid-infrared femtosecond high-repetition-rate master oscillator power amplifier (MOPA) based on a $Cr^{2+}$:ZnS gain medium. The MOPA delivers three-cycle pulses with a central wavelength of 2.35 μm at 80 MHz repetition rate, with maximum pulse energy 79 nJ. The output power of the MOPA is controlled by a half-wave plate combined with a germanium mirror placed at Brewster's angle. The reflected linearly-polarized beam from the germanium mirror is sent into the experiment vacuum chamber. The chamber is kept at a pressure of ~ $2 \times 10^{-2}$ mbar to avoid air absorption of ultraviolet radiation. The beam is focused onto the sample at normal incidence from the sapphire substrate side via an off-axis parabolic mirror with f/1.2, resulting in a beam diameter of 13 μm $1/e^2$ at the focus. We estimate that between ~60 to ~120 lines of the grating are illuminated within the beam diameter. A half-wave and a quarter-wave plate are placed before the chamber entrance to compensate the rotation of the polarization state of the excitation pulse in the birefringent sapphire substrate. The transmitted high harmonics are collected with a NA = 0.15 Fresnel zone plate (FZP) consisting of 0.5μm thick Si zones on a sapphire substrate. The FZP is fabricated in-house. The Si acts as an intensity mask for the harmonics, given that it is thicker than the absorption length of the harmonics reported. The FZP is designed to transmit the center portion of the high-harmonic beam (center-pass) to increase the transmissivity. The FZP focuses the harmonics onto the entrance slit of a VUV spectrometer (McPherson, model 234), equipped with a 1200 l/mm Al-coated grating. Given the low NA of the FZP compared to the estimated diffraction angle of the high harmonics from the gratings (> 20 degrees for the $13^{th}$ harmonic at the widest grating pitch), only the 0-th order should be collected by the FZP. The spectrally separated harmonics after the spectrometer are imaged with a Si CCD (Andor, model DO-420-BN). Due to slight chromatic dispersion of the FZP, its position is optimized for each harmonic order.

**Fabrication**

Grating structures are fabricated using a Si-on-sapphire wafer, with a nominal 1-μm-thick Si film and 460-μm-thick sapphire substrate. A JEOL JBX-6300FS electron-beam lithography system is used to pattern a 400-nm-thick FOX-16 electron-beam resist layer spun on Si surface. After e-beam lithography, a plasma etcher is used to transfer the pattern and etch through Si with $BCl_3/Cl_2/O_2$ chemistry. After etching, the exposed resist mask is stripped in HF.

The original design of the grating structure is 2 mm long and the width of the individual vacuum slot is linearly tapered from 50 nm to 500 nm over the total length. However, during electron-beam exposure, the narrow slot width side of the grating is over exposed due to proximity effect. The resist in the overexposed area forms a uniform oxide mask on Si and eventually leaves an area of unstructured Si film at the narrow side of the grating after etching, which is used to benchmark the bulk emission in the optical measurement.

After optical measurement, a few-nm thick of Au/Pd (60:40 ratio) layer is deposited on the structures for SEM imaging.

The same procedure is adopted to fabricate a center-pass Fresnel zone plate with 0.15 Numerical Aperture on a Si-on-sapphire 10x10 mm piece, with a nominal 500-nm-thick Si film and 460-μm-thick sapphire substrate.

**Data Availability** The datasets generated and/or analyzed during the current study are available from the corresponding author on reasonable request.

**Acknowledgments** This works is supported by the W. M. Keck Foundation and Stanford University. Part of this work was performed at the Stanford Nano Shared Facilities (SNSF), supported by the National Science Foundation (NSF) under award ECCS-1542152, and in the

nano@Stanford labs, which are supported by NSF as part of the National Nanotechnology Coordinated Infrastructure under award ECCS-1542152. The authors thank the Stanford Photonic Research Center and IPG Photonics for equipment gifts, R. Coffee for equipment loans, J. Vavra for equipment loans and discussions, and A. Sakdinawat and K. Lee for helping with the FZP fabrication.

**Author Contributions** H. L. and G. V. contributed equally. H. L. and G. V. designed the nanostructures and carried out the high-harmonic experiments. H. L. and J. L. Z. fabricated the structures. Y. S. and S. B. contributed to the simulations. J. V., S. F., P. H. B. and D. A. R. supervised the work. All authors contributed to the manuscript.

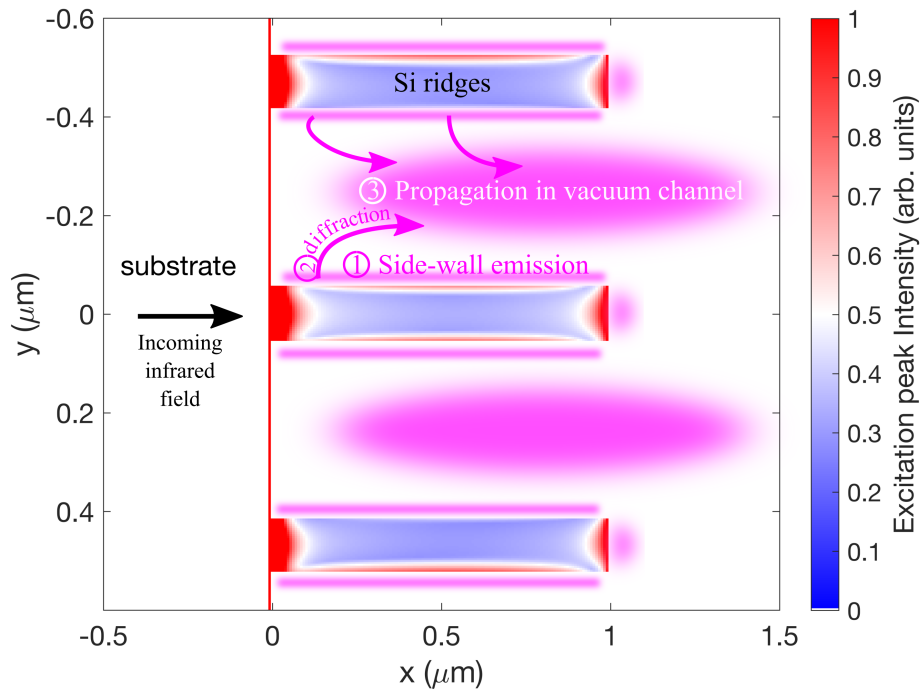

**Figure 1 | Working principle.** Cross-section of the designed grating structures composed of thin Si ridges, illuminated from left to right with an intense (~ 0.2 TW/cm$^2$ in vacuum) infrared laser field. The laser pulses propagate from left to right. The thin vertically red line represents the sapphire/Si interface. The infrared intensity simulated with Finite-Differences Frequency Domain (FDFD), color coded, is highest on the ridges' sidewalls, where high-harmonic generation preferentially takes place (step 1). Generation of high-harmonics within one

absorption length of bulk Si (~ 10 nm) from the Si-vacuum interface allows diffraction of the generated high-harmonics inside the vacuum slot (step 2), followed by their propagation and coherent addition in the slot (step 3). An illustration of the emitted high-harmonics is superimposed in purple.

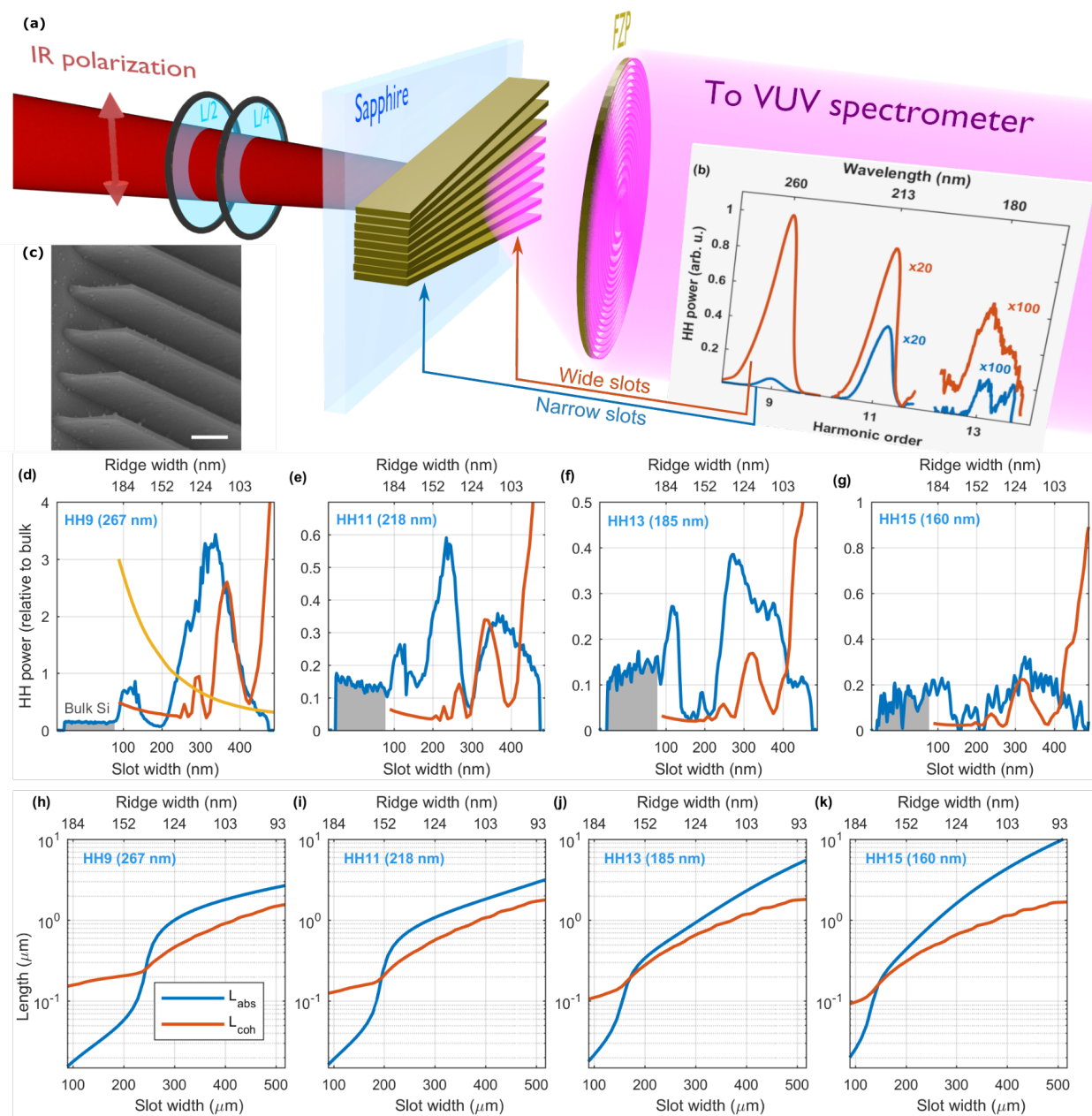

**Figure 2 | Experimental evidence. (a)** Sketch of the experimental apparatus. Linearly polarized mid-infrared femtosecond laser pulses (red beam) are focused on a chirped grating structure

composed of Si ridges with continuously varying width and pitch, fabricated on a sapphire substrate. Different geometries are explored by translating the grating laterally to the beam propagation axis. A combination of a half-wave plate (before) and a quarter-wave plate (after) are used to compensate for the birefringence of the sapphire substrate, assuring that the polarization entering the grating structure is linear and oriented Transverse Magnetic (TM). The generated high-harmonics (purple beam) are collected with a custom-made Fresnel Zone Plate (FZP) and focused on the input slit of a vacuum ultra-violet spectrometer. See methods for details on the experimental setup. **(b)** High-harmonic spectra recorded for two geometries: 180 nm slot width and 160 nm ridge width (blue line), 360 nm slot width and 110 nm ridge width (red line). **(c)** Scanning electron image of the end of the Si gratings. Scale bar is 500 nm (plane view). **(d-g)** Measured high-harmonic power (blue lines) for harmonics $9^{th}$ to $15^{th}$ (respectively **d** to **g**), as a function of the slot (bottom axis) and ridge (top axis) widths. The gray-shaded area corresponds to bulk Si. The flux is normalized to that obtained from bulk at the highest possible infrared power, just below the onset of bulk damage (see Fig. 3). The theoretical model (red lines) predicts an oscillating high-harmonic power with increasing amplitude. This power modulation occurs despite the monotonic decrease of the infrared field strength at the sidewalls of the Si ridges (yellow line in panel **d**, simulated with Finite-Differences Frequency Domain). **(h-k)** Calculated absorption and coherence lengths for the grating structures. For narrow slot widths, the short absorption length ($L_{abs}$, blue line) of the leaky slot mode limits the transmitted high-harmonic power to the intrinsic absorption limit of the bulk crystal (~ 10 nm). As the slot width widens beyond approximately one harmonic wavelength, transmission significantly increases compared to bulk. When the absorption length is larger than the coherent length ($L_{coh}$, red line), the phase mismatch over a distance of one coherence length, rather than material absorption, becomes the limiting factor.

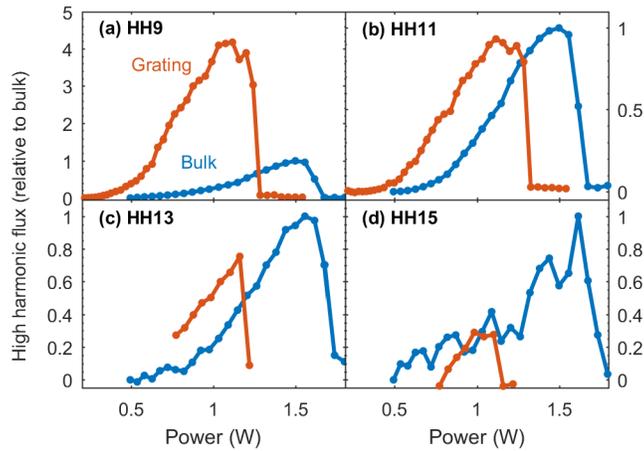

**Figure 3 | Power scaling.** The high-harmonic power from gratings (red lines) and bulk Si (blue lines) increases monotonically with the infrared power, regardless of the order, up to the onset of damage, marked by the sharp decrease. The red lines pertain the structures presented in Fig. 2. Each red curve is acquired at a geometry that maximizes the high-harmonic power in Fig. 2**d-g**.

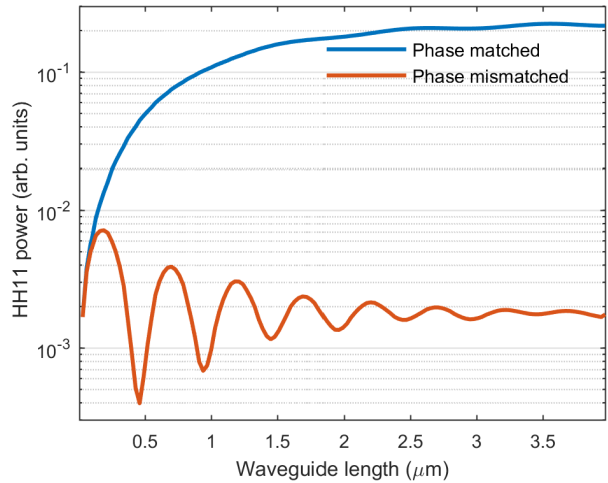

**Figure 4 | Phase-matching considerations.** As a result of the increased transmission for increasing slot widths, the predicted high-harmonic power oscillates with a period equal to one coherence length (red line). Equating the phase velocities of the harmonic and polarization waves may result in ~100-fold coherent build-up of the high-harmonic flux (blue line). Simulation is performed by varying the height of a grating, with fixed Si ridge width = 100 nm and vacuum slot width = 270 nm.

# Supplementary Information:

## Overcoming the absorption limit in high-harmonic generation from crystals

Hanzhe Liu[1,2*], Giulio Vampa[1*], Jingyuan Linda Zhang[3,4], Yu Shi[4,5], Siddharth Buddhiraju[4,5], Shanhui Fan[3,4,5], Jelena Vuckovic[3,4,5], Philip H. Bucksbaum[1,2,3,6], David A. Reis[1,3,6]

[1] *Stanford PULSE Institute, SLAC National Accelerator Laboratory, Menlo Park, California 94025, USA*

[2] *Department of Physics, Stanford University, Stanford, California 94305, USA*

[3] *Department of Applied Physics, Stanford University, Stanford, California 94305, USA*

[4] *E. L. Ginzton Laboratory, Stanford University, Stanford, California 94305, USA*

[5] *Department of Electrical Engineering, Stanford University, Stanford, California 94305, USA*

[6] *Department of Photon Science, Stanford University, Stanford, California 94305, USA*

\* These authors contributed equally to this work

**Simulation of the excitation field distribution**

We apply two-dimensional finite-difference-frequency-domain (FDFD) method to simulate the excitation field distribution in the grating structures. The total simulation region is a two-dimensional 8 μm × 2.5 μm grid and the spatial resolution along $x$ (horizontal) and $y$ (vertical) directions are set to be 4 nm and 7.8 nm, respectively. A Gaussian source, centered at 2.35 μm, is launched in $y$ direction one pixel below the Si grating in the sapphire substrate. The polarization of the source is set to be linear along $x$. We apply these simulation parameters to various Si gratings with different slot and ridge widths, matching the experiment conditions. The FDFD simulation outputs a complex field defined on the two-dimensional grid, containing both amplitude and phase information of the infrared field.

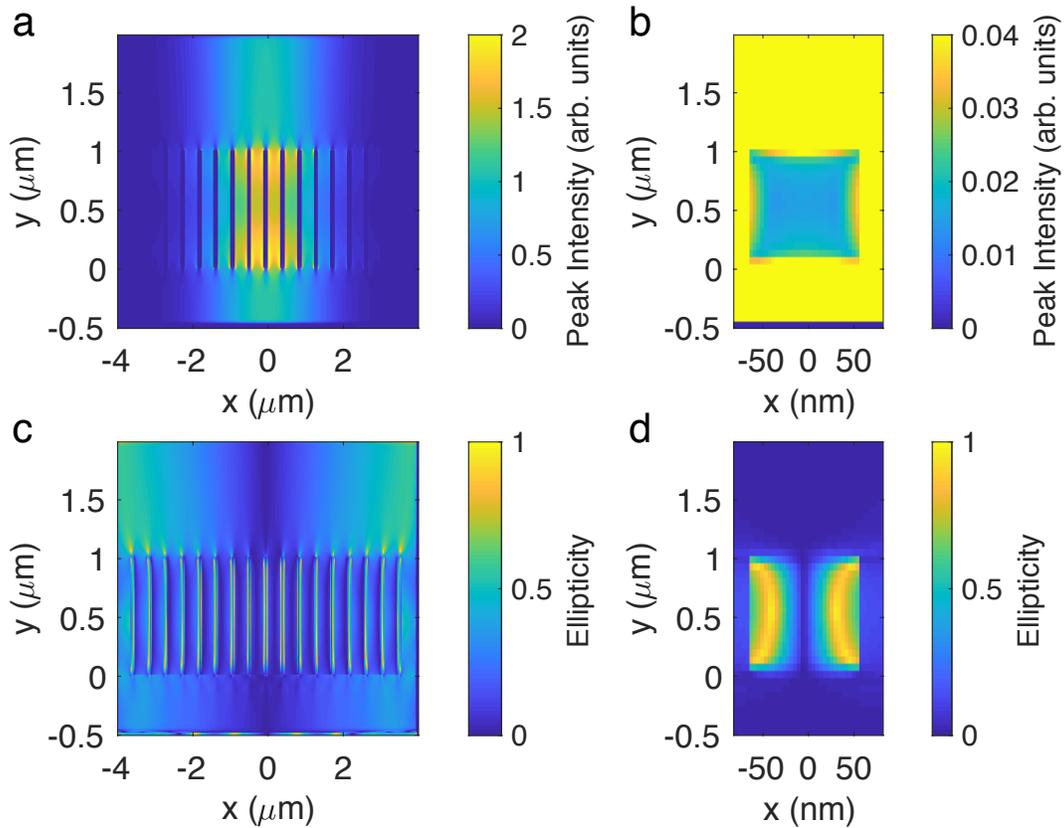

**Supplementary Figure 1 | Illustration of intensity and polarization of the excitation field in the grating structure at a specific geometry. a**, Excitation intensity distribution in the device. Each individual Si ridge is 1 μm tall and 118.4 nm wide. The width of the vacuum channel is 323.5 nm. **b**, Zoomed in image of intensity distribution inside a single Si ridge at the center of the grating. **c**, Ellipticity of the excitation field. **d**, Zoomed in image of ellipticity for the same Si ridge as in **b**.

Supplementary Figure 1**a** shows the intensity distribution of the infrared field in the grating structure for a specific geometry. Panel **b** represents a zoomed in plot for the Si ridge at the center of the grating structure. As the width of the individual Si ridge is extremely subwavelength, the excitation field on the side surface is higher than the field concentrated deep inside the Si, which is crucial for generating harmonics efficiently from the side surfaces and subsequently coupling them into the vacuum slot.

Even though the source is set to be linearly polarized along *x*, the field polarization in the grating structure is rotated and develops a non-zero component along *y*. The resulting ellipticity of the excitation field is illustrated in panel **c**. A closer examination of the ellipticity for the Si ridge at the center of the structure, as shown in panel **d**, reveals that the field is elliptically polarized along the sidewalls while linearly polarized at the top exit surface.

**Solving the high harmonic mode in the vacuum slot**

As the absorption length of the reported high harmonics in bulk Si is significantly smaller than the individual Si ridge width, each slot of the periodic grating can be considered independently. Thus, high harmonics can be considered propagating between the air gap of two infinitely extended parallel Si slabs, which constitutes a 1-dimensional leaky waveguide. The TM mode condition, in this case, is given by complex solutions of the following transcendental equation:

$$\tan(pa) = \frac{1}{n^2}\frac{q}{p}$$

where $p = \sqrt{k_0^2 - \beta_m^2}$, $q = \sqrt{\beta_m^2 - k_0^2 n^2}$, with $a, n, k_0, \beta_m$ being the vacuum slot half-width, Si refractive index at harmonic wavelength, harmonic wavevector in vacuum and the complex propagation constant, respectively. The complex refractive index of Si is obtained from Ref. [34]. The real and imaginary parts of the solution, $\beta_m$, correspond to propagation constant and absorption coefficient of the mode, respectively. The above equations are solved numerically.

**Simulation of phase mismatch effects**

In order to capture phase mismatch effects in the fabricated grating structures, we combine the FDFD simulation for the excitation field with the mode solution of harmonics in the vacuum slot. Specifically, we calculate the total high-harmonic power at the top exit of the grating. Assuming a grating with total height $L$, the harmonic emission generated from a distance $z$ below the top exit surface has the following contribution to the output:

$$E_m(z)e^{im\beta_0(L-z)}e^{iRe\{\beta_m\}z}e^{-\alpha z}$$

where $E_m(z), Re\{\beta_m\}, \beta_0, \alpha, m$ are local harmonic field amplitude, mode propagation constant of harmonics and excitation beam, absorption coefficient of the slot ($\alpha = Im\{\beta_m\}$) and harmonic order respectively. By virtue of the largely linear power scaling of high harmonics with the infrared power (Fig. 3), we approximate $E_m(z) = E_0(z)$, the local excitation field, which is obtained from the FDFD simulation. The first phase factor, $e^{im\beta_0(L-z)}$, is the phase of the nonlinear polarization at the harmonic wavelength. The second phase factor, $e^{i\beta_m z}$, is the propagation phase the harmonic accumulated from depth $z$ to the top exit surface. The exponential decay term, $e^{-\alpha z}$, characterizes the absorption of harmonics propagating in the vacuum slot. The total harmonic yield is given by:

$$I = \sum_{all\ slots} \left| \int_0^L dz\, E_0(z) e^{im\beta_0(L-z)} e^{i\beta_m z} e^{-\alpha z} \right|^2 + I_{top}$$

where $I_{top}$ is the harmonic intensity generated from the top surface of the Si grating. We apply this equation for grating structures with different geometry and the results are shown in Fig. 2 of the main text.

**Effects of infrared excitation on high-harmonic generation from the grating structure**

In this section we provide additional arguments that show that the observed order-dependent modulation of the high-harmonics reported in Fig. **2d-g** is not due to changes in the excitation infrared field.

As the slot width increases, both the excitation intensity and polarization on the sidewalls of the grating changes. However, as shown in Fig. 3, the high-harmonic flux increases (decreases) monotonically with the increasing (decreasing) infrared power for all harmonics simultaneously, up to the damage threshold (characterized by a sharp and irreversible drop of VUV flux). Therefore, a modulation of the infrared field strength with the waveguide geometry simultaneously changes all harmonics, contrary to the measured order-dependent modulation. Furthermore, the FDFD simulation predicts monotonically decreasing infrared field strength at the ridges' sidewalls as the slot widens, which cannot explain the non-trivial order-dependent modulation.

Polarization of the infrared field is also an important parameter in high-harmonic generation. When the pump pulse excites the TM mode of the grating structure, the excitation field in the structure develops a longitudinal component, resulting in elliptical polarization. This change of polarization state in the device can, in principle, locally alter the efficiency of high-harmonic emission. In this section, we characterize how the infrared polarization changes in the gratings and how this change affects high-harmonic generation. We find that, similarly to the field strength, the varying ellipticity and rotation of the infrared polarization is also expected to monotonically decrease high-harmonic emission as the slot widens. Thus, the observed harmonic modulation as a function of vacuum slot width is not due to the change in excitation fields.

Supplementary Figure 2 shows the ellipticity of the infrared field (panel **a**) and the ellipse's rotation (panel **b**), obtained from the FDFD simulation, as a function of slot and ridge widths. The geometries chosen in the simulation matches the fabricated ones. Both quantities are calculated at the point with maximum field strength along the Si ridge's sidewall. As the width of the vacuum slot increases, the ellipticity on the side surface increases from 0.4 to more than 0.7 in the reported slot width range, accompanied by a rotation of the field major axis. To quantify how this would impact harmonic emission, we measure the harmonic yield on an unpatterned Si film, with the Si film facing the incident beam, as a function of both excitation ellipticity and orientation of the field's major axis. Here, the same combination of half-wave and quarter-wave plate are used to control both ellipticity and rotation. The results are summarized in Supplementary Figure 3. For all three reported harmonics, harmonic emission is always more efficient when the field is linearly polarized. The generation process also exhibits a slight dependence on the excitation orientation, where the emission is maximized when the excitation is along (100) direction and minimized under (110) direction.

Under the experimental conditions reported in Fig. 2, as the vacuum slot width increases, the ellipticity of the excitation field on the sidewalls increases (shown in Supplementary Figure 1**a**) and its major axis rotates away from (100) crystallographic orientation (shown in Supplementary Figure 1**b**). Both effects will monotonically reduce the local high-harmonic emission efficiency and cannot explain the observed harmonic modulation as a function of the slot width, especially the sharp increase in harmonic yield when the slot width is comparable to the harmonic wavelength.

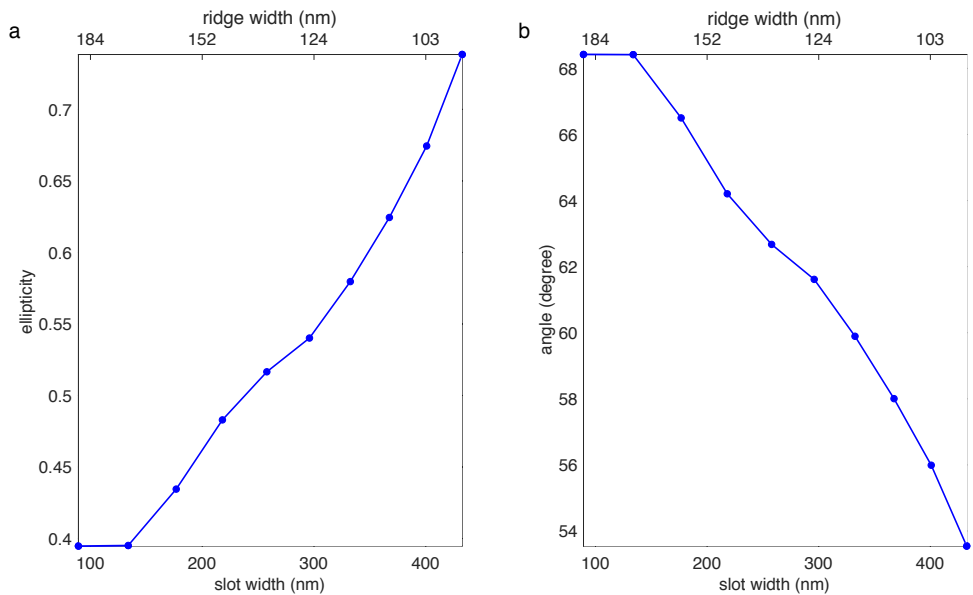

**Supplementary Figure 2 | Simulated ellipticity and ellipse's rotation along the Si ridge's sidewall**. For each slot width, ellipticity and rotation are calculated at the point of highest intensity along the sidewall length. (**a**) Infrared ellipticity as a function of slot width. (**b**) The orientation of the field major axis as a function of slot width, for the same points reported in **a**. The orientation is defined by the angle between the field major axis and the input polarization direction, which is TM in this case. 0 and 90 degree corresponds to (100) direction while 45 degree is along (110) direction.

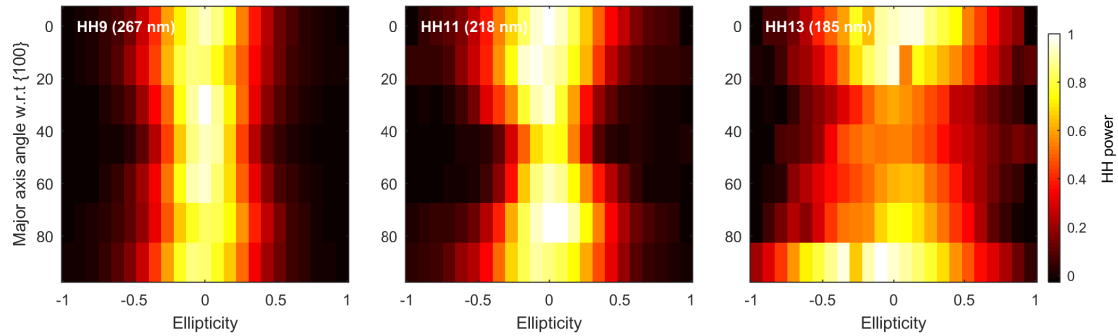

**Supplementary Figure 3 | Measured high-harmonic yield as a function of ellipticity and excitation orientation.** High-harmonic emission from bulk Si exhibits a much stronger dependence on ellipticity than rotation of the polarisation's major axis with respect to the crystallographic axis. High-harmonic emission is always maximized when the field is linearly polarized. For a given ellipticity, high-harmonic generation is maximized when the major axis of the field is along the (100) crystallographic direction and minimized along (110) direction.

**Slot-width dependence from thinner ridge gratings**

Supplementary Figure 4 reports the dataset equivalent to the one presented in Fig. 2**d-g**, for a grating structure with thinner Si ridges. Given that the pitch is identical to Fig. 2, a given ridge width corresponds to a wider slot width. Supplementary Figure 5 reports the high-harmonic scaling with the infrared power, similarly to Fig. 3, but including the scaling obtained from the thinner ridges of Supplementary Figure 4. Thinner ridges yield stronger emission than thicker ridges. In Supplementary Figure 5, the infrared power is varied at a geometry (slot and ridge widths) that yields the highest harmonic power in Supplementary Figure 4, for each harmonic separately.

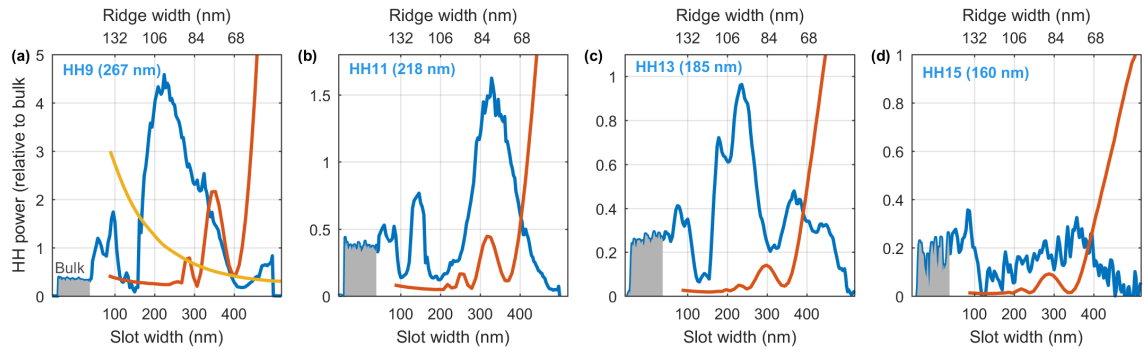

**Supplementary Figure 4** | Dependence of high-harmonic power on slot and ridge widths, for a grating with thinner ridges.

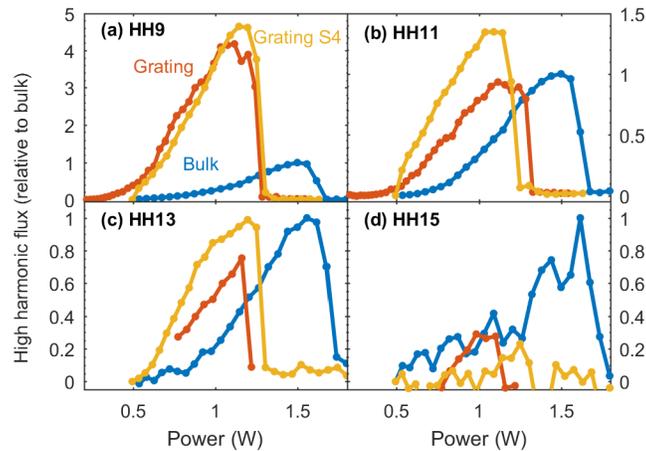

**Supplementary Figure 5** | Grating structures with the thinner Si ridges presented in this section yield higher maximum high-harmonic flux than those presented in Fig. 2. Here, the flux from the thinner structures as a function of infrared power (yellow lines, "Grating S4") is compared to that from bulk (blue lines, "Bulk", same dataset as in Fig. 3) and from the gratings presented in Fig. 2 (red lines, "Grating", same dataset as in Fig. 3). Each yellow curve is acquired at a geometry that maximizes the high-harmonic power in Supplementary Figure 4.

**References**

[34] www.refractiveindex.info